\documentclass[12pt]{iopart}
\usepackage{iopams}
\usepackage{graphics}
\begin{document}

\title[Thermodynamics of Evolving Lorentzian Wormholes in $f(R)$ Gravity]{Thermodynamics of Evolving Lorentzian Wormholes at Apparent Horizon in $f(R)$ Theory of Gravity}

\author{H. Saiedi } \ \\
\address{ }
\ead{\mailto {  hrssaiedi@gmail.com \ ; \   hrssaiedi@yahoo.com}}
\begin{abstract}\noindent \\
In the context of modified $f(R)$ gravity, we attempt to study the thermodynamic properties of the evolving Lorentzian wormholes at the apparent horizon.  It is shown that the wormhole can be derived from a particular $f(R)$ model in the radiation background.
Moreover, it has been shown that the field equations can be cast to a similar form $dE = TdS + WdV + Td\bar{S}$ at the apparent horizon for the evolving Lorentzian wormhole. Compared to the case of Einstein's general relativity, an additional term $Td\bar{S}$ appears here.

\noindent{\it Keywords\/}: modified gravity, evolving wormholes, apparent horizon, thermodynamics.
\end{abstract}
\pacs{04.50.-h, 04.50.Kd, 04.70.Dy}
%\maketitle

\section{Introduction}
The term wormhole  was first introduced by Flamm by means of the standard embedding diagram [1].  In 1988, the physics of wormhole was revived by the work of Morris and Thorne (MT). These authors proposed a possibility of traversable wormhole, through which observers can pass traveling  between two universes as a short cut [2]. According to the Einstein field equations, the MT wormhole needs the exotic matter (matter violating the null and weak energy conditions (NEC and WEC)) which holds up the wormhole structure and keeps the wormhole throat open. Recent astrophysical observations indicate that the universe could be dominated by a fluid which violates the null energy condition.

Although, an exotic matter is responsible for the early-time inflation and late-time acceleration (according to recent astrophysical data [3]), the gravitational alternative for a unified description of the inflation and dark energy is a very reasonable assumption. In other words, the modified gravity effectively describes the present acceleration and the early-time inflation without the need to introduce the exotic matter. Examples of such modified gravity models are theories where the Ricci scalar $R$ in the Einstein-Hilbert gravitational Lagrangian density is replaced by an arbitrary function $f(R)$. The literature on $f(R)$ gravity is vast, and the earlier reviews on this theory are given in [4, 5]. Different aspects of $f(R)$ gravity are discussed in [6, 7, 8].

The profound connection between gravity and
thermodynamics has been explored extensively. This connection was supported by the Stephen Hawking discovery that  black holes emit thermal radiation  corresponding to a temperature proportional to surface gravity and entropy proportional to the horizon area [9]. Therefore, there should be some relationship between thermodynamics and Einstein field equations. Jacobson is the first one who was able to derive Einstein equations from the proportionality of entropy to the horizon area of the black hole together with the first law of thermodynamics [10]. Padmanabhan made the major development by using  a general formalism for the spherically symmetric black hole spacetimes to understand the thermodynamics of horizons and showed that the Einstein field equations evaluated at event horizon can be expressed in the form of the first law of thermodynamics [11]. Later, Padmanabhan  and others studied this approach for more general spacetime geometries and in various gravity theories [12, 13].  This connection between gravity and thermodynamics has also been extended in the braneworld cosmology [14]. All these indicate that the thermal interpretation of gravity is to be generic, so we investigate this relation for a more general spacetimes. Hence, in this paper, we extend this approach for evolving Lorentzian wormhole spacetimes in $f(R)$ gravity and show that the field equations of the wormhole geometry can be expressed as the first law of thermodynamics at the apparent horizon.

It is interest to see how in the framework of $f(R)$ gravity, thermodynamic
properties of the wormhole spacetimes can be modified. Therefore, the idea that wormholes may show some characteristics and properties which are parallel to those already found in black holes, seems to be quite natural, including in particular "wormhole thermodynamics" [15]. In literature, the work on wormhole spacetimes in lower and higher dimensions have been studied by many authors [16]. For instance,  the discussion on Lorentzian wormholes in the n-dimensional Einstein gravity or Einstein-Gauss-Bonnet theory of gravitation is available in [17] while the author [18] discussed the interesting features of static wormhole in higher dimensional cosmology by considering the scale factor $a(t)$. Rahaman et al. [19] and Jamil and Farooq [20] constructed the phantom wormholes in the lower dimensional (2+1) gravity.

In this paper, we will deal with the modified $f(R)$ gravity. An evolving Lorentzian wormhole can be derived from a specific $f(R)$ gravity model in the radiation background. We will show that how the first law of thermodynamics is modified at the apparent horizon of the evolving wormhole.

\section{Evolving Lorentzian Wormhole in $f(R)$ Gravity}
A wormhole is a region of space-time with non-trivial topology. It has two mouths connected by a throat. The mouths are not hidden by event horizons as in the case of black holes, and, in addition there is no singularity to avoid the passage of particles from one side to the other. A simple generalization of MT wormhole to the time dependent background is given by the evolving Lorentzian wormhole [21] \\
\begin{equation}
ds^2 = - e^{2\Phi(t,r)} dt^2 + a^2(t) \left[ \frac{dr^2}{1 - \frac{b(r)}{r}
} +r^2 d\theta^2 + r^2 \sin^2 \theta d\phi^2 \right] \ ,
\end{equation} \ \\
where $\Phi(t,r), b(r)$ are the redshift and shape functions respectively and $a(t)$ is the scale factor. It is clear from the above metric that if $b(r)=Kr^3$ and $\Phi(t,r)=constant$ the metric reduces to the Friedmann-Robertson-Walker (FRW) metric. Moreover, by taking $a(t)=constant$ and $\Phi(t,r)=\Phi(r)$, it turns out the MT wormhole. In this paper we assume that the redshift function is zero,  $\Phi(t,r)=0$, without any loss of generality.

The action of modified $f(R)$ theory of gravity with the inclusion of matter is given by \\
\begin{equation}
A= \int {\sqrt{-g} \ (f(R) + 2L_m) \ d^4x} \ ,
\end{equation} \ \\
where $L_m$ \ is the matter Lagrangian density, $R$ is the  curvature scalar and $f(R)$  is an arbitrary function. We use the physical  units $8\pi G = c=1$ ( $G$ is Newton's gravitational constant and $c$ is the speed of light).  \\
The variational principle gives equations of motion \\
\begin{equation}
G_{\mu\nu} = \frac{1}{F}T^{m}_{\mu\nu} + \tilde{T}_{\mu\nu} = {T}^{e}_{\mu\nu} \ \ \ \ \ \ \ \ \ \ \mu,\nu=0, 1, 2, 3
\end{equation} \ \\
where $G_{\mu\nu}$ is the Einstein tensor, $F=df/dR \neq 0$ and \\
\begin{equation}
\tilde{T}_{\mu\nu} = T^{curv}_{\mu\nu} =  \frac{1}{F} \left [\frac{1}{2} g_{\mu\nu} \left ( f - R F \right) + \nabla_\mu\nabla_\nu F - g_{\mu\nu} \Box F \right ] ,
\end{equation} \ \\
and the stress-energy tensor $T^{m}_{\mu\nu}$ is written as \\
\begin{equation}
T^{\nu (m)}_{\mu} = diag ( - \rho(t,r), p_r(t,r), p_t(t,r), p_t(t,r) ) \ ,
\end{equation} \ \\
where $\rho(t,r), p_r(t,r)$ and $p_t(t,r)$ are the energy density, radial pressure and tangential pressure. \\
In the radiation background $(T = T^{\mu (m)}_{\mu} = - \rho + p_r + 2p_t = 0)$, by considering the specific shape function $b(r) = r_0$ and the scale factor $a(t) = a_0t^n \ (n>0)$,  the following exact solution for $f(R)$ can be achieved [22]. \\
\begin{eqnarray}
f(R)= -R \left[ c_1 \left( \frac{6n(2n-1)}{R} \right)^{\frac{m_{+}}{2}}   +   c_2 \left( \frac{6n(2n-1)}{R} \right)^{\frac{m_{-}}{2}} \right] \\ \nonumber \\ \nonumber
m_{\pm} = -\frac{3n}{2} + \frac{1}{2}  \pm  \frac{1}{2} \sqrt{57n^2-30n+1}
\end{eqnarray} \ \\
$c_1$ and $c_2$ are constant. \\
By substituting $b(r) = r_0$, the following relationships are obtained for the elements of the matter stress-energy tensor [22]. \\
\begin{eqnarray}
\rho = -\ddot{F} +3H^2F   \ , \\
p_r = - 2\dot{H}F + H\dot{F} -3H^2F - \frac{r_0 F}{a^2r^3} \ , \\
p_t = - 2\dot{H}F + H\dot{F} -3H^2F + \frac{r_0F}{2a^2r^3}  \ ,
\end{eqnarray} \ \\
where
\begin{equation}
F(t) = c_1 \ t^{m_{+}} + c_2 \ t^{m_{-}}
\end{equation} \ \\
and  $H=\dot{a}/a$, is the Hubble parameter. The overdot denotes differentiation with respect to time. \\
In the paper [22] author and Nasr Esfahani have shown that for $f(R)$ model (6), the matter threading the wormhole spacetimes  with either accelerated expansion or  decelerated expansion satisfies the NEC and WEC. It means that the evolving Lorentzian wormhole (1) with $b(r)=r_0, \ \ \Phi(t,r)=0$ and $a(t)=a_0t^n$ can be derived from the above $f(R)$ model in the radiation background which satisfies the null and weak energy conditions. In this section, we have reviewed the existence of evolving Lorentzian wormhole in the context of $f(R)$ gravity that authors  [22] have already discussed. In the next section we will deal with its thermodynamics.

\section{Wormhole Thermodynamics at Apparent Horizon}
It is important to note that the non-vanishing surface gravity at the wormhole throat characterized by a non-zero temperature for which one would expect that wormhole should emit some sort of thermal radiation. In this section, we discuss the thermodynamic properties of wormhole spacetimes at the apparent horizon within the $f(R)$ gravity model (6). For this purpose, one can rewrite the metric (1) with $b(r)=r_0$ and $ \Phi(t,r)=0$ in the spherical form  \\
\begin{equation}
ds^2 = h_{ab} dx^a dx^b + \tilde{r}^2 (d\theta^2 +  \sin^2 \theta d\phi^2 )  \ \ \ \ , \ \ \ \ a,b = 0, 1
\end{equation} \ \\
where $\tilde{r}=a(t)r , x^0 = t , x^1 = r$ and the two dimensional metric $h_{ab}$ is written as \\
\begin{equation}
h_{ab} = diag  \left[-1 ,  a^2(t) \left( 1 - \frac{r_0}{r}     \right)^{-1} \right] \ .
\end{equation} \ \\
The dynamical apparent horizon can be evaluated by using the relation $h^{ab} \partial_a\tilde{r} \partial_b\tilde{r} = 0 $, which after simplification we have \\
\begin{equation}
H^2 \tilde{r}_A^3 - \tilde{r}_A + r_0 a(t) = 0 \ ,
\end{equation} \ \\
where $\tilde{r}_A$ represents the apparent horizon radius of the evolving wormhole. It can be seen from (13) that at $r_0 = 0$, namely a flat FRW universe, the wormhole apparent horizon has the same value as the Hubble horizon. The Hubble parameter in terms of the wormhole apparent radius is $H^2 = 1/\tilde{r}_A^2 - ar_0/\tilde{r}_A^3$, and its time derivative is given as
\begin{equation}
\dot{H} = \frac{\dot{\tilde{r}}_A}{2H\tilde{r}_A^3} \left( \frac{3ar_0}{\tilde{r}_A} - 2 \right) - \frac{ar_0}{2\tilde{r}_A^3} \ .
\end{equation} \ \\

Now, by combining equations (7), (8) we reach the following equation  \\
\begin{equation}
\dot{H}  + \frac{ar_0}{2\tilde{r}^3}       = - \frac{1}{2F} \left[ (\rho + p_r) + \ddot{F} - H\dot{F} \right] \ .
\end{equation} \ \\
Substituting the equation (14) into the above equation, one can obtain the following differential equation for the apparent horizon. \\
\begin{equation}
- H \left[ (\rho + p_r) + \ddot{F} - H\dot{F} \right] dt = \frac{F}{\tilde{r}_A^3} \left( \frac{3ar_0}{\tilde{r}_A} - 2 \right) d\tilde{r}_A
\end{equation} \ \\
Since the area of the wormhole horizon is $A = 4\pi \tilde{r}_A^2$, so one can relate the entropy with the surface area of the apparent horizon through $S = \frac{A}{4G}\mid_{\tilde{r}_A}$. In the $f(R)$ theory of gravity, the entropy has the following relation to the horizon area \\
\begin{equation}
S = \frac{AF}{4G}\mid_{\tilde{r}_A} =  \frac{\pi \tilde{r}_A^2 F}{G} = 8\pi^2 \tilde{r}_A^2 F \ \ \ \ \ \ \ \ \ \ \ (8\pi G = 1) \ ,
\end{equation} \ \\
so that
\begin{equation}
dS = 8\pi^2 \tilde{r}_A^2 dF + 16\pi^2 \tilde{r}_AF d\tilde{r}_A \ .
\end{equation} \ \\
By using the above relation, the equation (16) can be written as \\
\begin{eqnarray}
\frac{1}{2\pi\tilde{r}_A} \left( 1 - \frac{3ar_0}{2\tilde{r}_A} \right) dS &=& 4\pi\tilde{r}_A^3 H \left[ (\rho + p_r) + \ddot{F} - H\dot{F} \right] dt \nonumber \\ &+& 4\pi\tilde{r}_A \left( 1 - \frac{3ar_0}{2\tilde{r}_A} \right) dF \ \ .
\end{eqnarray} \ \\

The surface gravity is defined as [23] \\
\begin{equation}
\kappa = \frac{1}{2\sqrt{-h}} \ \partial_a \left( \sqrt{-h} \  h^{ab} \partial_b \tilde{r}   \right) \ ,
\end{equation} \ \\
where $h$ is the determinant of metric $h_{ab}$  (12). The direct calculation of the surface gravity from (20) at the wormhole horizon $\tilde{r}_A$ yields \\
\begin{eqnarray}
\kappa &=& - \frac{\tilde{r}_A}{2} \left( \dot{H} + 2 H^2 - \frac{ar_0}{2\tilde{r}_A^3}   \right) \nonumber \\
&=& - \frac{1}{\tilde{r}_A}  \left( 1 - \frac{\dot{\tilde{r}}_A}{2H\tilde{r}_A}   \right) \left( 1 - \frac{3ar_0}{2\tilde{r}_A}   \right) \ .
\end{eqnarray} \ \\
The factor $ - \frac{1}{\tilde{r}_A}  \left( 1 - \frac{\dot{\tilde{r}}_A}{2H\tilde{r}_A}   \right) $ is the general expression for the surface gravity of FRW universe while the second factor $\left( 1 - \frac{3ar_0}{2\tilde{r}_A}   \right)$ has been appeared due to the wormhole geometry. In the case of  $r_0 = 0$, the expression for the surface gravity (21) reduces to the expression for FRW universe. The apparent horizon temperature is defined as $T = \kappa /2\pi$. In order to have a positive temperature, we should use $T = |\kappa| /2\pi$ to define the temperature of the apparent horizon. Now, our next step is to write the differential equation (19) as a first law of thermodynamics for the evolving Lorentzian wormhole in $f(R)$ gravity model which mentioned in the previous section. \\
By multiplying both sides of the differential equation (19) by a factor $- \left( 1 - \dot{\tilde{r}}_A / 2H\tilde{r}_A  \right) $ and arranging the terms, one can obtain \\
\begin{eqnarray}
T dS =  &-& 4\pi\tilde{r}_A^3 H (\rho + p_r) \left( 1 - \frac{\dot{\tilde{r}}_A}{2H\tilde{r}_A}   \right) dt   \nonumber \\
&-&  4\pi\tilde{r}_A^3 H (\ddot{F} - H\dot{F} ) \left( 1 - \frac{\dot{\tilde{r}}_A}{2H\tilde{r}_A}   \right) dt  \ + \  8\pi^2 \tilde{r}_A^2 T dF \ .
\end{eqnarray} \ \\

The unified first law is defined by [24] \\
\begin{equation}
dE = A\Psi + WdV
\end{equation} \ \\
where  $A=4\pi\tilde{r}^2$, $V=\frac{4}{3}\pi\tilde{r}^3$ and $E$ are the area,  volume and total energy of matter, respectively. The unified first law expresses the gradient of the active gravitational energy divided into energy-supply and work terms. The energy-supply (the first term on the right hand side of (23)) on the horizon is the total energy flow through the apparent horizon, while the work (the second term) at the apparent horizon should be regarded as the work done by a change of the apparent horizon. \\
The work density $W$ and energy-supply vector $\Psi_a$ are given by \\
\begin{equation}
W = - \frac{1}{2} h^{ab}T_{ab} = \frac{1}{2} (\rho - p_r)  \ \ \ \ , \ \ \ \ \Psi_a = T_a^b\partial_b\tilde{r} + W \partial_a\tilde{r}
\end{equation} \ \\
and we know that $\Psi = \Psi_a dx^a$. Therefore, by using the above relations and after some simplification we reach the following relation at the apparent horizon. \\
\begin{equation}
A\Psi = 2\pi\tilde{r}_A^2 ( \rho + p_r )d\tilde{r}_A - 4\pi\tilde{r}_A^3H ( \rho + p_r )dt
\end{equation} \ \\
One can immediately find that the first term on the right hand side of the equation (22) can be rewritten as the equation (25). Hence, we rewrite  (22) as \\
\begin{equation}
T dS = A\Psi - T d\bar{S}
\end{equation} \ \\
where
\begin{equation}
d\bar{S} = - 8\pi^2\tilde{r}_A^2 \left[ \dot{F} + \frac{\tilde{r}_A^2}{ \left(1-\frac{3ar_0}{2\tilde{r}_A} \right)} H (\ddot{F} - H\dot{F} )  \right] dt \ \ .
\end{equation} \ \\
By using (23), the equation (26) becomes  \\
\begin{equation}
dE = T dS + W dV + T d\bar{S} \ .
\end{equation} \ \\
In the cosmological setting, it has been shown that the field equations obey the universal form $dE = T dS + W dV$ at the apparent horizon of evolving wormholes in Einstein gravity [23]. Here, we see from the equation (28) that the form $dE = T dS + W dV$ does not hold at the apparent horizon of evolving Lorentzian wormholes in $f(R)$ gravity. One can choose $f(R) = R$, in this case the additional term $Td\bar{S}$ will be vanished and the equation (28) will be turned out to the case of Einstein gravity.
If $d\tilde{S} = d(S + \bar{S})$ is defined as an effective entropy change during the infinitesimal displacement $d\tilde{r}_A$ of the apparent horizon of the evolving Lorentzian wormhole in interval $dt$, the equation (28) can be rewritten as \\
\begin{equation}
dE = T d\tilde{S} + W dV \ .
\end{equation} \ \\
where $\tilde{S} = S + \bar{S}$ is the effective entropy associated with the apparent horizon of the evolving Lorentzian wormhole in $f(R)$ theory of gravity. Now, we can see how  the  thermodynamic properties of the evolving Lorentzian wormhole spacetimes can be modified in the context  of modified  $f(R)$ gravity,

\section{Conclusion}
In this work, we have reviewed the evolving Lorentzian wormhole spacetimes in $f(R)$ gravity in the radiation background. We impose that the matter threading the wormhole satisfies the null and weak energy conditions. However, the field equations of gravity in Einstein, Gauss-Bonnet and Lovelock gravity can be cast to a similar form of the first law of thermodynamics, $dE = TdS + WdV$, at the apparent horizon of the evolving wormhole, in $f(R)$ gravity it is different. Here, $E$ is the total energy of matter inside the apparent horizon, $T$ and $S$ are the associated temperature and entropy with the horizon, $W$ and $V$ are the work density and volume inside the horizon, respectively. It has been shown that the field equations of the wormhole geometry in $f(R)$ gravity can be rewritten as $dE = TdS + WdV + Td\bar{S}$ at the apparent horizon. The additional term $d\bar{S}$ can be regarded as an entropy production term associated with the apparent horizon in the non-equilibrium thermodynamics within the $f(R)$ gravity. It will be interesting in further investigation to study the  thermodynamic properties of other sorts of wormhole spacetimes in various gravity theories.

\end{document}